\documentclass[preprint,12pt]{elsarticle}

\usepackage{graphicx}
\usepackage{subfigure}
\usepackage{amsmath,amssymb,bm}
\usepackage{gensymb}

\usepackage{lineno}

\usepackage{color,soul}

\journal{Applied Energy}

\begin{document}

\begin{frontmatter}


\title{Optimal Utilization Strategy of the LiFePO$_4$ Battery Storage}



\author{Timur Sayfutdinov\fnref{label1,label2}\corref{cor1}}
\author{Petr Vorobev\fnref{label1}}

\address[label1]{Center for Energy Science and Technology, Skolkovo Institute of Science and Technology, 3 Nobel Street, Skolkovo, Moscow Region 121205, Russia}
\address[label2]{School of Engineering, Newcastle University, Newcastle upon Tyne, United Kingdom}
\cortext[cor1]{t.saifutdinov@newcastle.ac.uk}

\begin{abstract}
The paper provides a comprehensive battery storage modelling approach, which accounts for operation- and degradation-aware characteristics and can be used in optimization problem formulations. 
Particularly, Mixed-Integer Linear Programming (MILP) compatible models \textcolor{black}{have been developed for} the lithium iron phosphate (LiFePO$_4$) battery storage \textcolor{black}{using the \emph{Special Order Sets 2} to represent the nonlinear characteristics, including} efficiency, internal resistance growth, and capacity fade. 
Such formulation can be used in problems related to various applications, i.e., power systems, smart grid, and vehicular applications, and it allows finding the globally optimal solution using off-the-shelf academic and commercial solvers. 
In the numerical study, the proposed modelling approach has been applied to realistic scenarios of peak-shaving, where the importance of considering the developed models is explicitly demonstrated. 
\textcolor{black}{Operation- and degradation-aware techno-economic analysis showed that the optimal battery capacity is driven by operating rather than service requirements. Particularly, a considerable battery over-sizing becomes economically feasible when the battery storage is used more extensively. Another finding suggests that to achieve the maximum value from battery storage, its operation strategy needs to be significantly modified during the course of its lifetime. In the scenarios considered, the charging time gradually increased from four to seven hours, while the average SoC decreased by 20\%.}
Such an adaptable scheduling results in reduced battery degradation and a longer lifetime, which may provide \textcolor{black}{as much as} 12.1\% of savings in the battery storage system project.
\end{abstract}


\end{frontmatter}


\section{Introduction}\label{sec:introduction}
Nowadays, energy storage systems have established their efficacy for more than a dozen power system applications, which cover all stages in the energy supply chain: bulk power and energy; ancillary services; transmission and distribution infrastructure applications; customer energy management \cite{akhil2013doe}. In its turn, the electrification of transport heavily relies on the developments in storage technologies. \textcolor{black}{Among} all storage technologies used in power systems and transportation, lithium-ion (Li-ion) batteries are the fastest-growing \cite{navigant2019}. In either application, optimal (or near-optimal) utilization of batteries is ensured by thorough planning and scheduling. Notably, online scheduling of battery storage is implemented by an energy management system, which produces optimal dispatch signals based on the operation requirements and current battery state, where the latter is estimated using various online functions, e.g., battery cells' consistency evaluation \cite{wang2020novel}, battery aging assessment \cite{she2019battery}, and open-circuit voltage reconstruction \cite{tian2021electrode}. In its turn, the planning process implies performing a techno-economic analysis prior to battery installation, which is usually done employing various optimization methods \cite{zidar2016review}. The result of such an analysis is typically an optimal choice for storage unit siting, sizing, and technology selection, as well as the optimal charge/discharge scheduling, i.e., operation strategy. To ensure informed decision making, both planning and scheduling tasks require operation- and degradation-aware battery modelling, which is provided in the present paper.

In early optimization problem formulations, such as in \cite{dvorkin2016ensuring,xu2013optimal}, constant efficiency for charge and discharge were considered when modelling battery behavior. In practice, efficiency is a function of the battery output current and also the battery state parameters, which include internal resistance and open-circuit voltage, that change significantly with the battery State of Charge (SoC), temperature, and State of Health (SoH) \cite{plett2015battery}. For instance, it was shown in \cite{alvaro2020} that charge and discharge efficiencies may vary significantly - they can drop as much as $33\%$ from their maximum values depending on the battery operating conditions. To account for the influence of power output and SoC on battery efficiency, \cite{morstyn2017model} proposed a second-order polynomial formulation, which can be considered within the convex programming approach. Then, a Mixed-Integer Linear Programming (MILP) compatible representation of the Li-ion battery has been proposed in \cite{sakti2017enhanced}, where efficiency was modelled using a piece-wise linear approximation of the simulated sample data. As an efficient alternative, \cite{alvaro2020} proposed a Linear Programming (LP) framework to account for efficiency based on the equivalent circuit model while still considering the MILP formulation in \cite{sakti2017enhanced} as a benchmark.

While focusing on a more accurate representation of battery efficiency, the above-mentioned references did not account for an \emph{operation-aware lifetime} and, most importantly, for the \emph{available energy capacity} of the Li-ion battery storage, which decreases gradually over its lifetime due to degradation. The very first attempts to represent operation-aware battery lifetime were models based on the total energy throughput, as in \cite{sayfutdinov2018incorporating}. To respect the nonlinear relationship between battery operation strategy, i.e., Depth of Discharge (DoD), and its cycle lifetime, \cite{alsaidan2018comprehensive} approximated the dependency using a piece-wise linear formulation and considered it within a MILP framework for optimal battery sizing problem. Next, in \cite{padmanabhan2019battery} previous approaches were enhanced by incorporating $C$-rate as an additional factor of battery wear-and-tear. However, the methods above did not account for the inevitable capacity loss of Li-ion batteries over their lifetime, which plays one of the most important roles in the techno-economic analysis of battery storage.

Extensive experimental studies in the literature, considered in the next section, suggest that the battery degradation depends in a more complicated (often nonlinear) way on a number of factors, such as battery SoC, temperature, DoD, etc. Thus, certain approximations have to be made to account for these effects when formulating an optimization problem for techno-economical analysis. In early attempt \cite{qiu2017stochastic}, a constant capacity fade rate of Li-ion battery was introduced for the storage investment problem. Even though the degradation rate was considered to be fixed, irrespective of the battery operation, the results suggest that capacity fade is among the most important factors to account for. In addition to the previous effect, in \cite{miranda2016holistic,parra2015optimum} the battery available capacity was considered to be fading in time proportionally to the energy throughput. Considering the degradation rate to be dependant on operation variables, i.e., battery power output, made the optimization problem bilinear and required applying the whole enumeration search to find the globally optimal solution. In \cite{sayfutdinov2019degradation}, dynamic programming and mixed-integer problem reformulation approaches have been proposed to consider operation-aware degradation from SoC and DoD while still respecting the formal optimization requirements. In \cite{maheshwari2020optimizing}, the short-term operation strategy of the Li-ion battery storage has been investigated using the MILP problem formulation, where the nonlinear cycling degradation effect from SoC, DoD, and C-rate has been captured using the piece-wise linear approximation. In \cite{li2020design,berrueta2018combined}, comprehensive Li-ion battery models were formulated for the optimal sizing problem, where the capacity fade effect from both idling and cycling mechanisms were complemented with the phenomenon known as the internal resistance growth, which affects the battery maximum power output and efficiency. Both models are characterized with the nonlinear formulation, which were approached with two distinct methods. Particularly, the particle swarm optimization heuristic has been used in \cite{li2020design}, while a formal approach of dynamic programming has been applied in \cite{berrueta2018combined}, where the former method cannot guarantee the optimality of a solution and the latter possesses a high computational burden.

\textcolor{black}{The literature review suggests that the state-of-the-art methods for techno-economic evaluation of the Li-ion batteries are proposed in \cite{li2020design} and \cite{berrueta2018combined}. Similar to the state-of-the-art, the proposed approach comprehensively accounts for all Li-ion battery effects considered in \cite{li2020design} and \cite{berrueta2018combined}. However, in contrast to the state-of-the-art, the proposed modelling approach exploits formal optimization method, i.e., MILP, that allows finding the globally optimal solution in a computationally efficient way by applying various partial enumeration techniques (e.g., branch-and-bound). Moreover, the proposed MILP compatible model extends the state-of-the-art by the capability to find optimal operation strategies for each individual battery lifetime period (e.g., year), while in the literature, a single operation strategy is found for the whole battery lifespan resulting in a sub-optimal solution.}

In the proposed modelling approach, the lithium iron phosphate (LiFePO$_4$) battery model is developed based on the existing experimental literature. The model includes the realistic \textcolor{black}{nonlinear} dependencies of \textcolor{black}{operational characteristics (i.e., operation strategy) on efficiency, lifetime, and available capacity, which were linearized} using the \emph{Special Order Sets 2}. Then the formulation of an optimization problem \textcolor{black}{has been proposed} for the optimal choice of battery size and operation strategy for realistic case studies, where the operation strategy can be adjusted for each battery lifetime period individually, i.e., optimization problem variables. \textcolor{black}{The proposed MILP compatible model can be applied for both optimal planning and scheduling of battery storage, which can provide real value to battery storage owners (e.g., transmission and distribution network operators) by performing a comprehensive degradation- and operation-aware techno-economic analysis. Notably, the considered LiFePO$_4$ battery possesses all characteristics attributed to Li-ion technologies (i.e., capacity fade, internal resistance growth, SoC dependent internal resistance, and open-circuit voltage), making the proposed MILP formulation suitable for other chemistries (e.g., lithium manganese oxide, lithium nickel manganese cobalt oxide, lithium titanium oxide).}

\textcolor{black}{The numerical study suggests} that there exist a number of trade-offs when deciding on a particular battery size and operation strategy, where the former might be significantly bigger than the minimum required capacity, and the latter should be modified over the whole battery lifetime to provide economically optimal result. Particularly, to achieve optimal utilization of the LiFePO$_4$ battery, its capacity may exceed the minimum service requirement by \textcolor{black}{as much as} 77.3\%, its average SoC needs to be altered by up to 20\%, while the duration of the charging process is required to increase by up to 75\% during the battery lifetime. Compared to the state-of-the-art methodology, the associated economic effect of the proposed approach accounts for a 12.1\% of reduction of battery investment and operating costs. Even though the proposed approach has been demonstrated for the LiFePO$_4$ battery, the methodology is applicable to other types of Li-ion family of technologies. 

To summarize, the main contributions of the present manuscript are the following:
\begin{enumerate}
\item A MILP compatible Li-ion battery model\textcolor{black}{, which} is based on the experimental results and accounts for realistic operation-aware efficiency and degradation, including capacity fade and internal resistance growth.
\item \textcolor{black}{Optimal LiFePO$_4$ battery operation strategies obtained for real cases of peak-shaving that require significant life-long modifications to
achieve the maximum value from the battery use.}
\item \textcolor{black}{Identified trade-offs in the LiFePO$_4$ battery that vary over the lifetime and alternately affect the optimal battery operation strategy during different time periods. For instance, in the early battery lifetime, capacity fade from cycling is minimized at the cost of the increased degradation from idling, while, closer to the terminal year, increased losses (due to the internal resistance growth) are compensated at the cost of increased capacity fade from idling.}
\end{enumerate}

The rest of the paper is organized as follows: 
Section 2 presents Li-ion battery models, including storage continuity model, equivalent circuit model, and battery degradation models;
Section 3 provides MILP compatible problem formulation for the optimal scheduling and sizing of the Li-ion battery storage;
Section 4 presents results of the numerical study for peak-shaving application;
Section 5 provides information on the software and hardware used and engages discussion on the applicability of the proposed method;
conclusions and future research are drawn in Section 6.

\section{Li-ion Battery Modelling}\label{sec:model}

In the present study, a MILP compatible LiFePO$_4$ battery model \textcolor{black}{is built} based on the available experimental data from the literature provided in the present section. Particularly, the efficiency model \textcolor{black}{is formulated} based on the equivalent circuit Li-ion battery models from \cite{greenleaf2014application,morstyn2017scalable}. The open-circuit voltage dependency for the LiFePO4 battery is taken from \cite{plett2015battery}. The internal battery resistance characteristic of the LiFePO$_4$ battery from SoC and the number of equivalent full cycles is taken from \cite{wang2017online}. The LiFePO$_4$ battery capacity fade characteristics from cycling and idling have been taken from \cite{stroe2016degradation}.

The central part in energy storage modelling is a storage continuity differential equation, which tracks the battery charge. In a general form, it looks as follows
\begin{equation}\label{eq:charge_diff}
    \dot{e}=P^\text{B},
\end{equation}
where $e$ is a battery charge and $P^\text{B}$ is a battery power input. While the former cannot take negative values, the latter is positive when the battery charges and negative when it discharges.

The battery power input $P^\text{B}$ accounts for the amount of power drawn in and out of the battery cells. Due to power losses present in real cells, the battery power input $P^\text{B}$ is different from the power seen at terminals $P^\text{T}$ - power that goes to/from the grid. In the most simplistic representation, the ratio of $P^\text{B}$ and $P^\text{T}$ is considered to be constant, which corresponds to constant battery efficiency. In reality, the efficiency depends on the battery operation parameters as well as on its SoH. In the present study, the equivalent circuit representation \textcolor{black}{is used} to approximate the relationship between $P^\text{B}$ and $P^\text{T}$.

\subsection{Equivalent circuit model}

\begin{figure}
\centering
        \includegraphics[width=0.4\textwidth]{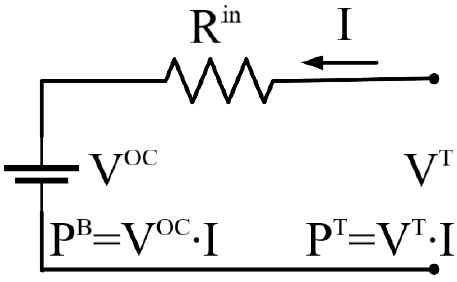}
        \caption{Rint model of a Li-ion cell}
        \label{fig:rint}
\end{figure}

Equivalent circuit modelling is an efficient tool to represent complex phenomena using circuit theory. A comprehensive electric circuit model for Li-ion cells derived from a physics-based phenomenological model has been provided in \cite{greenleaf2014application}. The model incorporates a number of $RLC$ circuits connected in series that represent dynamics of electrochemical processes, and it is mainly used for dynamic studies. However, due to nonlinearity, such a detailed model is found to be intractable for optimization tasks. In fact, this degree of detail is found to be redundant for the applications where the time-scale is significantly longer than the transient time constant, i.e., scheduling and sizing. Thus, given the fact that the aggregate time constant of transient processes of Li-ion batteries is in the order of minutes \cite{plett2015battery}, a steady-state model can be effectively used for the optimal siting and scheduling problems, where the characteristic time-scale is of the order of hours or half-hours. The equivalent steady-state model would corresponds to a circuit that contains voltage source and effective series resistance as depicted in Fig. \ref{fig:rint} - $Rint$ model \cite{morstyn2017scalable}.

Given the $Rint$ model of Fig. \ref{fig:rint}, the battery power input $P^\text{B}$ can be expressed as a function of the power at terminals $P^\text{T}$ and battery state parameters, i.e., open-circuit voltage $V^\text{OC}$ and internal resistance $R^\text{in}$,
\begin{equation}\label{eq:Pb}
    P^\text{B}=\frac{V^\text{OC}\sqrt{{V^\text{OC}}^2 + 4 P^\text{T} R^\text{in}} - {V^\text{OC}}^2}{2 R^\text{in}}.
\end{equation}

\begin{figure}
\centering
        \includegraphics[width=0.6\textwidth]{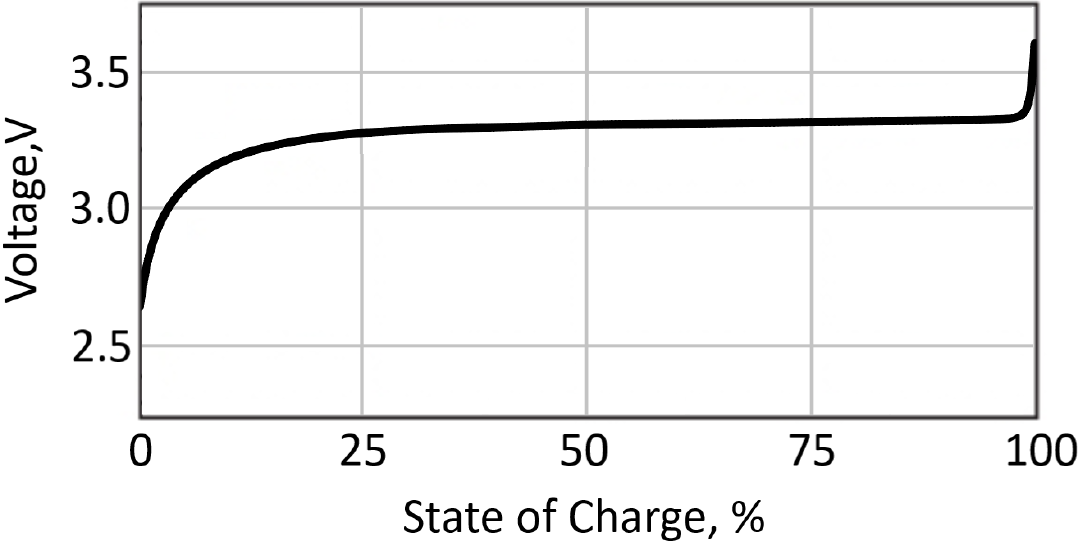}
        \caption{LiFePO$_4$ battery open-circuit voltage vs SoC}
        \label{fig:ocv_soc}
\end{figure}

The first element of the $Rint$ model is a voltage source, with voltage level $V^\text{OC}$ dependent on the battery SoC. Fig. \ref{fig:ocv_soc} illustrates the dependency of the LiFePO$_4$ battery open-circuit voltage and SoC state value at 25\degree C \cite{plett2015battery}. For Li-ion chemistries, the dependency is considered to be linear within a wide range of SoC. Particularly, for LiFePO$_4$ batteries, it is found to be linear between 10\% and 98\% SoC. Thus, it can be effectively approximated using the following linear relation:
\begin{equation}\label{eq:Voc}
    V^\text{OC}(SoC)=\text{k}^\text{V} SoC + V^\text{0},
\end{equation}
where $\text{k}^\text{V}$ is a voltage slope and $V^\text{0}$ is an offset value, e.g., for LiFePO$_4$ battery $\text{k}^\text{V}=0.15$ V/pu, $V^\text{0}=3.2$ V.

\begin{figure}
\centering
        \includegraphics[width=0.5\textwidth]{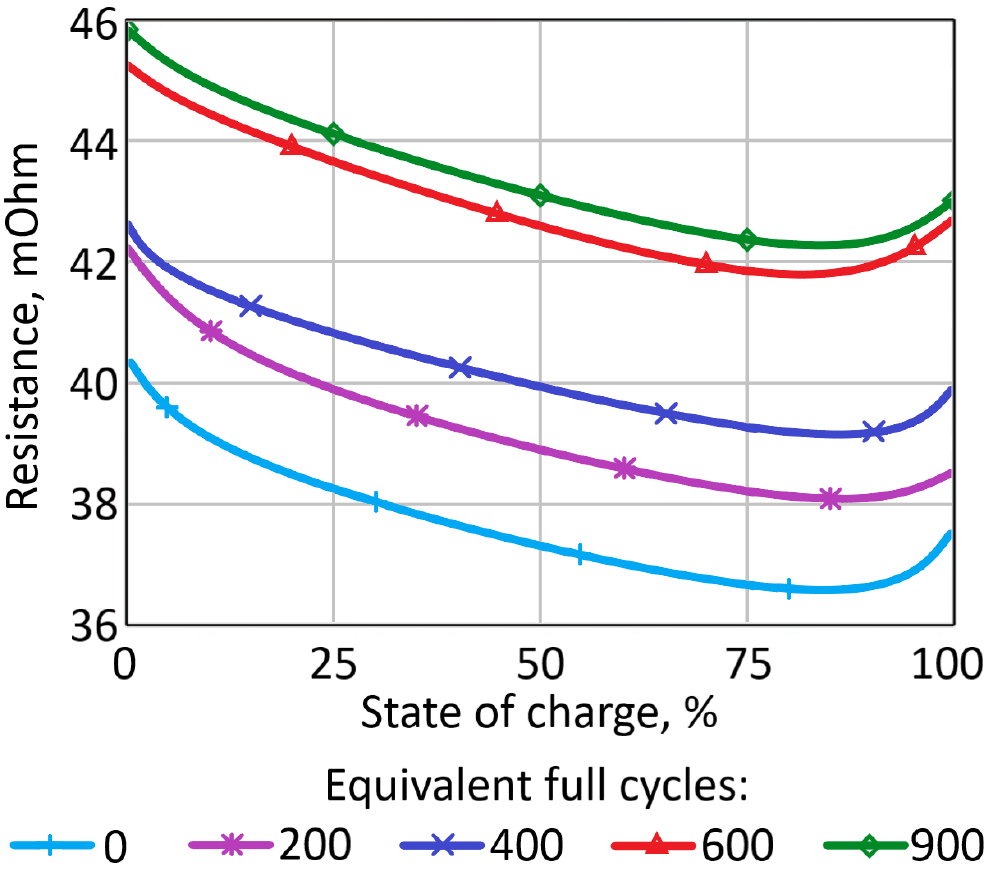}
        \caption{Resistance vs SoC}
        \label{fig:Rin}
\end{figure}

The second element of the $Rint$ model is the internal resistance $R^\text{in}$, which incorporates a series of resistive elements of the original model \cite{greenleaf2014application} and depend on the state of the battery, including SoC and SoH, where the latter sometimes is expressed in the equivalent full cycles. Fig. \ref{fig:Rin} illustrates the relationship of the internal battery resistance from SoC and the number of equivalent full cycles at 25\degree C \cite{wang2017online}. The figure shows that for all aging states, the internal resistance value is nearly linear for the wide range of SoC [10\%;85\%], while a substantial increase is observed near the minimum and maximum SoC. Such non-monotonous function can be effectively linearized using the three linear segments, which represent the internal resistance characteristics for low, medium, and high ranges of SoC. At the same time, the value of internal resistance increases monotonously with the equivalent full cycles and can be approximated with a single linear function. Thus, the battery internal resistance can be represented with the combination of linear functions as follows:
\begin{equation}\label{eq:Rin}
    R^\text{in}= \text{a}_k^\text{SoC}SoC_k + \text{b}_k^\text{SoC} + \text{a}^\text{FC}N^\text{FC},
\end{equation}
where \textcolor{black}{$k$ denotes a particular linear segment, $SoC_k$ is SoC value within the $k$-th linear segment (i.e., $\text{SoC}_k$ range), $\text{a}_k^\text{SoC}$ and $\text{b}_k^\text{SoC}$ are linear and constant coefficients of the characteristic from the SoC value, while $\text{a}^{\text{FC}}$ is a linear term of the dependency from the number of equivalent full cycles $N^\text{FC}$. Table~1 provides piecewise-linear function coefficients, where the particular values have been obtained by fitting the linear function \eqref{eq:Rin} to the empirically obtained data from \cite{wang2017online} (illustrated in Fig.~\ref{fig:Rin}) using the least-squares method \cite{matlabcurvefitting}.}


\begin{table}
\caption{\textcolor{black}{Piecewise-linear function coefficients of the internal resistance characteristic}}\label{Table:rin}
\begin{center}
\begin{tabular}{c | c | c | c | c}
$k$ & \text{SoC}$_{k}$ range & $\text{a}_{k}^{\text{SoC}}$, \textit{mOhm/pu} & $\text{b}_{k}^{\text{SoC}}$, \textit{mOhm} & $\text{a}^{\text{FC}}$, \textit{mOhm/cyc} \\ \hline
    1        & 0\% - 10\%         & -13.3         & 40.39 & \\
    2        & 10\% - 85\%        & -3.44         & 39.44 & 0.0064\\
    3        & 85\% - 100\%       & 6.72          & 30.86 & \\
\hline
\end{tabular}
\end{center}
\end{table}

\begin{figure}
\centering
        \includegraphics[width=0.75\textwidth]{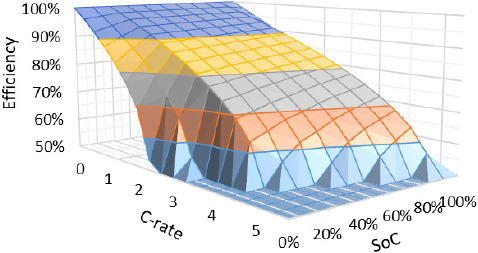}
        \caption{Discharge efficiency vs SoC and C-rate}
        \label{fig:efficiency_3d}
\end{figure}

To estimate the losses obtained by the proposed $Rint$ model and the dependencies above, the charge and discharge efficiencies can be found as a ratio between $P^\text{B}$ and $P^\text{T}$, depending on the power flow direction. Fig. \ref{fig:efficiency_3d} illustrates battery discharge efficiencies derived from (\ref{eq:Pb}) for RCR123A 0.45Ah LiFePO$_4$ cell from \cite{greenleaf2014application} at the beginning of its lifetime. It can be noted that even at a moderate discharge rate of 1C, one-way efficiency may drop below 90\%.

\subsection{Degradation model}

From the operational perspective, the most important aspects of the Li-ion battery degradation are internal resistance growth and capacity fade. While the former influences the maximum power output and losses, the latter affects the available energy capacity during the battery lifetime. 

The battery internal resistance growth is associated with the Solid Electrolyte Interface (SEI) formation on the surface of the anode \cite{rodrigues1999ac}. The SEI resistance increases with every cycle through the whole battery lifetime, which is considered by the second term in (\ref{eq:Rin}). As reported in \cite{wang2017online}, the total internal resistance increases nearly linearly with the number of equivalent full cycles, rising by as much as 20\% per $1,000$ full cycles.

The next aspect of the battery degradation is a continuous decrease of available capacity - capacity fade. There are two main degradation mechanisms considered in the literature, namely, idling $\delta^\text{idl}$ and cycling $\delta^\text{cyc}$, and the total capacity loss $\delta^\text{CF}$ can be approximated as a sum of both contributions \cite{schmalstieg2014holistic}: 
\begin{equation}\label{eq:cap_fade}
    \delta^\text{CF} \approx \delta^\text{idl}+\delta^\text{cyc}.
\end{equation}

Degradation from cycling implies that the available capacity decreases after each charge-discharge cycle, and the amount of the capacity loss is driven by the charge and discharge rate (C-rate), cycle DoD and SoC range, and cell temperature during the cycle \cite{stroe2016degradation}. At the same time, idling degradation implies that the available capacity is lost, even when the battery is not being cycled. The rate of capacity fade in this case depends on the state of the battery, i.e., SoC and cell temperature \cite{grolleau2014calendar}. In \cite{stroe2016degradation}, empirical capacity fade models due to both idling and cycling are provided based on the accelerated aging tests results:
\begin{equation}\label{eq:cap_fade_idl}
    \delta^\text{idl}= 0.000112 e^{0.7388 SoC^\text{idl}} \tau^{0.8},
\end{equation}
\begin{equation}\label{eq:cap_fade_cyc}
    \delta^\text{cyc}= 0.00568 e^{-1.943 SoC^\text{cyc}} DoD^{0.7162} \sqrt{n},
\end{equation}
where $SoC^\text{idl}$ is the average battery SoC, $\tau$ is time in days, $SoC^\text{cyc}$ is the SoC level around which a cycle is made, i.e., median cycle SoC, $DoD$ is the cycle DoD, and $n$ is the number of cycles. \textcolor{black}{The particular values of model parameters have been obtained in \cite{stroe2016degradation} by fitting the empirically obtained data to the model developed using the least-squares method.}

It can be noted that both (\ref{eq:cap_fade_idl}) and (\ref{eq:cap_fade_cyc}) are formulated for the cell temperature of $25$\degree C, which is considered to be constant in \textcolor{black}{the present} study. \textcolor{black}{Similar to the state-of-the-art methods in \cite{li2020design} and \cite{berrueta2018combined}, it is assumed that the battery storage system is equipped with a thermal management system that, in combination with a low C-rate power output corresponding to most battery storage applications, results in a flat temperature profile of battery storage \cite{patsios2016integrated}.}

\begin{figure}
\centering
        \includegraphics[width=0.75\textwidth]{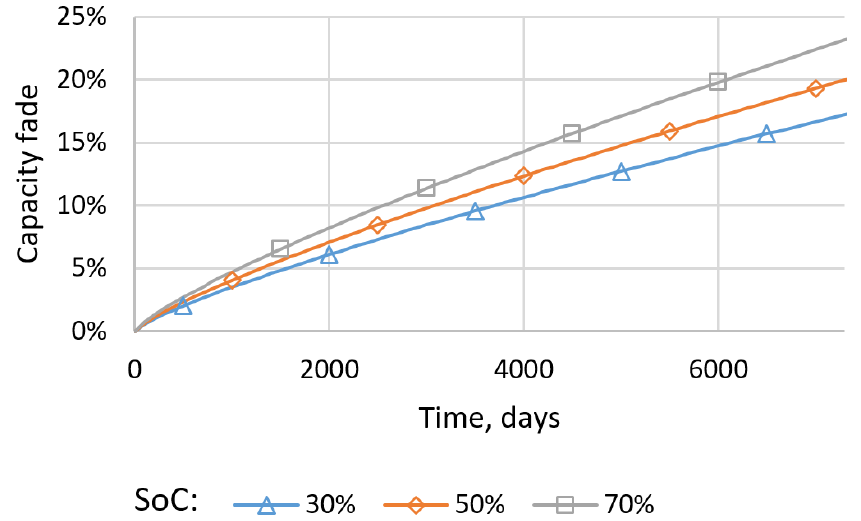}
        \caption{Capacity fade due to idling}
        \label{fig:cap_fade_idl}
\end{figure}

\begin{figure}
\centering
        \includegraphics[width=0.75\textwidth]{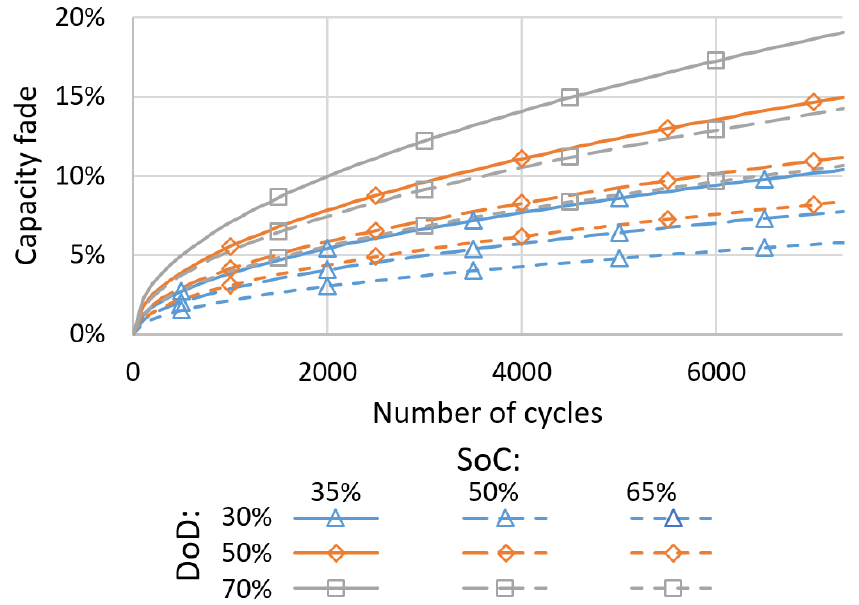}
        \caption{Capacity fade due to cycling}
        \label{fig:cap_fade_cyc}
\end{figure}

\textcolor{black}{For illustrative purposes, Fig.~\ref{fig:cap_fade_idl} and Fig.~\ref{fig:cap_fade_cyc} depict the capacity fade characteristics \eqref{eq:cap_fade_idl} and \eqref{eq:cap_fade_cyc} obtained in \cite{stroe2016degradation} from the accelerated aging tests.} 
Particularly, Fig.~\ref{fig:cap_fade_idl} illustrates that capacity fade from idling \textcolor{black}{\eqref{eq:cap_fade_idl}} is slower when the battery SoC is kept low. From this figure, one can infer that it is in general better to keep the battery discharged when the service is not required. On the other hand, Fig.~\ref{fig:cap_fade_cyc} suggests that capacity loss from cycling \textcolor{black}{\eqref{eq:cap_fade_cyc}} is the most severe for high DoD and low median SoC. Thus, to decrease capacity loss from cycling, one would want to charge and discharge the battery around the highest possible SoC. Obviously, the above degradation mechanisms disagree and require a balanced trade-off to ensure optimal battery utilization.



\section{Optimization Problem Formulation}\label{sec:case_study}

In the present section, a generic optimization problem \textcolor{black}{is formulated} for the techno-economic analysis of battery storage deployment. The proposed formulation is in the class of Mixed-Integer Linear Programming (MILP) problems, where the objective function and constraints are linearly formulated with respect to continuous and integer variables. The particular formulation allows using off-the-shelf numerical algorithms (e.g., the interior-point algorithm for linear terms and branch-and-bound for integer terms) to find the global minimum or maximum of an objective function while ensuring that the constraints are satisfied.

The particular problem below is looking at the optimal scheduling and sizing of the Li-ion battery storage, where the battery is aimed to deliver power according to predetermined demand profiles, e.g., peak-shaving and EV drive cycle. In this formulation, the battery follows dynamics specified in the constraints of the optimization problem (i.e., battery models) to deliver predetermined discharge power, while the battery charging schedule, installed power and energy capacities, as well as the idling and cycling characteristics (i.e., average SoC, median cycle SoC, and cycle DoD), are being optimized to deliver the service in the most economical way. The corresponding objective function is formulated as follows
\begin{equation}\label{eq:objective}
    min{[\frac{\text{C}^\text{E} \bar{\text{E}} + \text{C}^\text{P} \bar{P}}{365 \: \text{T}^\text{Lt}} + \sum_{y\in Y}\pi_y \sum_{t\in T}(-\text{C}^\text{LL} P_{y,t}^\text{LL} + \text{C}^\text{En} (P_{y,t}^{\text{T}_\text{Ch}} + P_{y,t}^{\text{T}_\text{Dis}}))\Delta t}],
\end{equation}
where $Y$ is a set of operation scenarios (e.g., years) indexed by $y$, $T$ is a set of time intervals indexed by $t$ with a time step $\Delta t$. $\pi_y$ is a normalized probability of a scenario $y$. $\bar{\text{E}}$ and $\bar{P}$ are installed energy and power capacities of the battery, $\text{C}^\text{E}$ and $\text{C}^\text{P}$ are the corresponding prices for the installed capacities, which all together make the investment cost of energy storage. To consider the investment cost in the same time-scale as the daily operating costs, the later is divided by the battery lifetime $365 \: \text{T}^\text{Lt}$, which also corresponds to a planning horizon. The battery power input at terminals is broken into positive charge $P_{y,t}^{\text{T}_\text{Ch}}$ and negative discharge $P_{y,t}^{\text{T}_\text{Dis}}$ to avoid nonlinear problem formulation. $\text{C}^\text{En}$ is a price for energy, necessary to translate power losses accounted in (\ref{eq:Pb}) into pecunial losses. $P_{y,t}^\text{LL}$ is a slack variable to allow minor deviations from the power balance equality (\ref{eq:balance}), which is penalized by the value of lost load $\text{C}^\text{LL}$.

To ensure that the battery delivers power according to predetermined demand profiles, the following  power balance and thermal line limit constraints are applied
\begin{align}
    P_{y,t}^\text{G} + \text{P}_{y,t}^\text{D} + P_{y,t}^{\text{T}_\text{Ch}} + P_{y,t}^{\text{T}_\text{Dis}} & \: + P_{y,t}^\text{LL}= 0 \label{eq:balance},\\
    -\bar{\text{P}}^\text{G} \leq P_{y,t}^\text{G} \leq& \: 0 \label{eq:thermal_limit},
\end{align}
where $P_{y,t}^\text{G}$ is a power supplied from the grid, $\bar{\text{P}}^\text{G}$ is the line thermal limit and $\text{P}_{y,t}^\text{D}$ is a power demand profile.

To model battery storage, the linear and mixed-integer linear constraints are formulated below. First, storage continuity differential equation (\ref{eq:charge_diff}) in a discrete form looks as follow
\begin{equation}\label{eq:SOC}
    e_{y,t+1}=(1-\text{k}^\text{sd})e_{y,t} + (P_{y,t}^{\text{B}_\text{Ch}} + P_{y,t}^{\text{B}_\text{Dis}}){\Delta t},
\end{equation}
where $\text{k}^\text{sd}$ is a self-discharge rate and battery power input $P^\text{B}$ from (\ref{eq:Pb}) is broken into positive charge $P_{y,t}^{\text{B}_\text{Ch}}$ and negative discharge $P_{y,t}^{\text{B}_\text{Dis}}$ to avoid nonlinear problem formulation.

Net storage charge, power rating, available storage capacity and maximum capacity fade are respected through (\ref{eq:net_charge}) - (\ref{eq:eol_cond})
\begin{align}
    e_{y,1} = e_{y,\text{T}+1}& \label{eq:net_charge},\\
    0 \leq P_{y,t}^{\text{T}_\text{Ch}} \leq \bar{P} & \label{eq:ch_limit},\\
    -\bar{P} \leq P_{y,t}^{\text{T}_\text{Dis}} \leq 0 & \label{eq:dis_limit},\\
    0 \leq e_{y,t} \leq \bar{\text{E}}(1 - & \: \delta_y^\text{CF}) \label{eq:charge_limit},\\
    \delta_y^\text{CF} \leq 1 - \text{EoL} & \label{eq:eol_cond},
\end{align}
where $\delta_y^\text{CF}$ is a battery capacity fade and $\text{EoL}$ is End of Life criterion, i.e., minimum remaining battery capacity threshold.

Before approximating nonlinear battery power input and capacity fade using \emph{Special Order Sets 2} it is required that the reference variables are broken into segments as in (\ref{eq:pt_ch})-(\ref{eq:soc_cur})
\begin{equation}\label{eq:pt_ch}
    P_{y,t}^{\text{T}_\text{Ch}} = \sum_{g=1}^{\text{G}}P_{y,t,g}^{\text{T}_\text{Ch}},
\end{equation}
\begin{equation}\label{eq:pt_dis}
    P_{y,t}^{\text{T}_\text{Dis}} = \sum_{h=1}^\text{H}P_{y,t,h}^{\text{T}_\text{Dis}},
\end{equation}
\begin{equation}\label{eq:dod_cyc}
    \frac{1}{2 \bar{\text{E}}}\sum_{t \in T_c}(P_{y,t}^{\text{B}_\text{Ch}} - P_{y,t}^{\text{B}_\text{Dis}}) \Delta t = \sum_{i=1}^\text{I}DoD_{y,c,i}^\text{cyc}
\end{equation}
\begin{equation}\label{eq:soc_cyc}
    \frac{\min_{t \in T_c}\{e_{y,t}\}}{\bar{\text{E}}} + \frac{\sum_{i=1}^\text{I}DoD_{y,c,i}^\text{cyc}}{2} = \sum_{l=1}^\text{L}SoC_{y,c,l}^\text{cyc},
\end{equation}
\begin{equation}\label{eq:soc_idl}
    \frac{1}{\bar{\text{E}} \text{T}}\sum_{t \in T} e_{y,t} \Delta t = \sum_{j=1}^\text{J}SoC_{y,j}^\text{idl},
\end{equation}
\begin{equation}\label{eq:soc_cur}
    \frac{e_{y,t}}{\bar{\text{E}}} = \sum_{k=1}^\text{K}SoC_{y,t,k},
\end{equation}
where segmented $P_{y,t,g}^{\text{T}_\text{Ch}}$ and $P_{y,t,h}^{\text{T}_\text{Dis}}$ are charge and discharge power outputs, $DoD_{y,c,i}^\text{cyc}$ and $SoC_{y,c,l}^\text{cyc}$ are cycle DoD and median SoC, $SoC_{y,j}^\text{idl}$ is the average daily SoC and $SoC_{y,t,k}$ is momentary SoC. $T_c$ is a time range of a cycle $c$, $\text{T}_c$ is a cycle duration and $\text{G}, \text{H}, \text{I}, \text{J}, \text{K}, \text{L}$ are the numbers of segments. In (\ref{eq:soc_cyc}), the minimum battery charge during a cycle is found with the following reformulation
\begin{equation}\label{eq:e_min}
    \min_{t \in T_c}\{e_{y,t}\} = e_{y,c}^\text{min},
\end{equation}
\begin{equation}\label{eq:e_min_con}
    e_{y,c}^\text{min} \leq e_{y,t} \: \forall t \in T_c.
\end{equation}

To ensure that the segments in (\ref{eq:pt_ch})-(\ref{eq:soc_cur}) are filled in the consecutive manner, the following constraints are applied
\begin{align}
    |\text{P}_g^{\text{T}_\text{Ch}}| \alpha_{y,t,g+1} \leq P_{y,t,g}^{\text{T}_\text{Ch}} \leq |\text{P}_g^{\text{T}_\text{Ch}}| \alpha_{y,t,g},g=1..\text{G} \label{eq:pt_ch_aux},\:\:\:\:\:\: \\
    |\text{P}_h^{\text{T}_\text{Dis}}| \beta_{y,t,h+1} \leq P_{y,t,h}^{\text{T}_\text{Dis}} \leq |\text{P}_h^{\text{T}_\text{Dis}}| \beta_{y,t,h} ,h=1..\text{H} \label{eq:pt_dis_aux},\:\:\:\:\: \\
    |\text{DoD}_i^\text{cyc}| \gamma_{y,c,i+1} \leq DoD_{y,c,i}^\text{cyc} \leq |\text{DoD}_i^\text{cyc}| \gamma_{y,c,i},i=1..\text{I} \label{eq:dod_cyc_aux},\\
    |\text{SoC}_l^\text{cyc}| \zeta_{y,c,l+1} \leq SoC_{y,c,l}^\text{cyc} \leq |\text{SoC}_l^\text{cyc}| \zeta_{y,c,l},l=1..\text{L} \label{eq:soc_cyc_aux},\:\: \\
    |\text{SoC}_j^\text{idl}| \eta_{y,j+1} \leq SoC_{y,j}^\text{idl} \leq |\text{SoC}_j^\text{idl}| \eta_{y,j},j=1..\text{J} \label{eq:soc_idl_aux},\:\:\:\:\: \\
    |\text{SoC}_k| \theta_{y,t,k+1} \leq SoC_{y,t,k} \leq |\text{SoC}_k| \theta_{y,t,k},k=1..\text{K} \label{eq:soc_cur_aux}, \:\:
\end{align}
where $\alpha_{y,t,g}, \beta_{y,t,h}, \gamma_{y,c,i}, \zeta_{y,c,l}, \eta_{y,j}, \theta_{y,t,k}$ are auxiliary binary variables, which indicate if a particular segment is used, and the binaries for the indices $\text{G}+1, \text{H}+1, \text{I}+1, \text{J}+1, \text{K}+1, \text{L}+1$ are enforced to zeros and considered as parameters. Finally, $|\cdot|$ is a length of a particular segment.

Now capacity fade can be approximated as follows
\begin{multline}\label{eq:cap_fade_lin}
    \delta_{y+1}^\text{CF} = \delta_{y}^\text{CF} + \sum_{c=1}^\text{C} [ \sum_{i=1}^\text{I}(\gamma_{y,c,i}-\gamma_{y,c,i+1}) \sum_{l=1}^\text{L}(\zeta_{y,c,l}-\zeta_{y,c,l+1}) \\ \cdot \frac{\partial \delta^\text{cyc}(\hat{\text{DoD}}_{y,c,i}^\text{cyc}, \hat{\text{SoC}}_{y,c,l}^\text{cyc}, 365\text{C}(y-0.5))}{\partial n}365] + \\ + \sum_{j=1}^\text{J}(\eta_{y,j}-\eta_{y,j+1}) \frac{\partial \delta^\text{idl}(\hat{\text{SoC}}_{y,j}^\text{idl},365(y-0.5))}{\partial \tau}365,
\end{multline}
where $\text{C}$ is a number of cycles performed during a scenario. The partial derivatives of capacity fade from idling (\ref{eq:cap_fade_idl}) and cycling (\ref{eq:cap_fade_cyc}) are found for the corresponding lifetime moments, i.e, time, number of performed cycles, cycle DoD $\hat{\text{DoD}}_{y,c,i}^\text{cyc}$, median cycle SoC $\hat{\text{SoC}}_{y,c,l}^\text{cyc}$ and average daily SoC $\hat{\text{SoC}}_{y,j}^\text{idl}$. The latter three are found as follows
\begin{equation}\label{eq:dod_cyc_str}
    \hat{\text{DoD}}_{y,c,i}^\text{cyc} = \sum_{i'=1}^\text{i-1}|\text{DoD}_{i'}^\text{cyc}| + \frac{|\text{DoD}_{i}^\text{cyc}|}{2},
\end{equation}
\begin{equation}\label{eq:soc_cyc_str}
    \hat{\text{SoC}}_{y,c,l}^\text{cyc} = \sum_{l'=1}^\text{l-1}|\text{SoC}_{l'}^\text{cyc}| + \frac{|\text{SoC}_l^\text{cyc}|}{2},
\end{equation}
\begin{equation}\label{eq:soc_idl_str}
    \hat{\text{SoC}}_{y,j}^\text{idl} = \sum_{j'=1}^\text{j-1}|\text{SoC}_{j'}^\text{idl}| + \frac{|\text{SoC}_j^\text{idl}|}{2}.
\end{equation}
The product of binary variables in (\ref{eq:cap_fade_lin}) is substituted with a variable $u_{y,c,i,l} = \gamma_{y,c,i} \zeta_{y,c,l}$, which is linearized as in (\ref{eq:bin_prod1_lin})
\begin{equation}\label{eq:bin_prod1_lin}
    \begin{gathered}
    0 \leq u_{y,c,i,l} \leq 1,\\
    u_{y,c,i,l} \leq \gamma_{y,c,i},\\
    u_{y,c,i,l} \leq \zeta_{y,c,l},\\
    u_{y,c,i,l} \geq \gamma_{y,c,i} + \zeta_{y,c,l} - 1.
    \end{gathered}
\end{equation}

Next, charge and discharge battery power output is approximated as follows
\begin{multline}\label{eq:Pb_ch}
    P_{y,t}^{\text{B}_\text{Ch}} = \prod_{y'=1}^y \sum_{i=1}^\text{I}(\gamma_{y',c,i} - \gamma_{y',c,i+1}) \sum_{k=1}^\text{K}(\theta_{y,t,k} - \theta_{y,t,k+1}) \\ \sum_{g=1}^\text{G} \frac{\partial P^\text{B}(\hat{\text{P}}_g^{\text{T}_\text{Ch}},\hat{\text{SoC}}_{y,t,k},\hat{\text{N}}_{I(y)}^{\text{FC}})}{\partial P^\text{T}} P_{y,t,g}^{\text{T}_\text{Ch}}
\end{multline}
\begin{multline}\label{eq:Pb_dis}
    P_{y,t}^{\text{B}_\text{Dis}} = \prod_{y'=1}^y \sum_{i=1}^\text{I}(\gamma_{y',c,i} - \gamma_{y',c,i+1}) \sum_{k=1}^\text{K}(\theta_{y,t,k} - \theta_{y,t,k+1}) \\ \sum_{h=1}^\text{H} \frac{\partial P^\text{B}(-\hat{\text{P}}_h^{\text{T}_\text{Dis}},\hat{\text{SoC}}_{y,t,k},\hat{\text{N}}_{I(y)}^{\text{FC}})}{\partial P^\text{T}} P_{y,t,h}^{\text{T}_\text{Dis}}
\end{multline}
where the partial derivative of the battery power output (\ref{eq:Pb}) is found for each segment of terminal power outputs $\hat{\text{P}}_g^{\text{T}_\text{Ch}}$ and $\hat{\text{P}}_h^{\text{T}_\text{Dis}}$, momentary SoC $\hat{\text{SoC}}_{y,t,k}$ and the number of full equivalent cycles $\hat{\text{N}}_{I(y)}^{\text{FC}}$, which are found as follows
\begin{equation}\label{eq:Pb_ch_aux1}
    \hat{\text{P}}_g^{\text{T}_\text{Ch}} = \sum_{g'=1}^{g-1}|\text{P}_{g'}^{\text{T}_\text{Ch}}| + \frac{|\text{P}_{g}^{\text{T}_\text{Ch}}|}{2},
\end{equation}
\begin{equation}\label{eq:Pb_dis_aux1}
    \hat{\text{P}}_h^{\text{T}_\text{Dis}} = \sum_{h'=1}^{h-1}|\text{P}_{h'}^{\text{T}_\text{Dis}}| + \frac{|\text{P}_{h}^{\text{T}_\text{Dis}}|}{2},
\end{equation}
\begin{equation}\label{eq:Pb_ch_aux2}
   \hat{\text{SoC}}_{y,t,k} = \sum_{k'=1}^{k-1}|\text{SoC}_{k'}| + \frac{|\text{SoC}_{k}|}{2},
\end{equation}
\begin{equation}\label{eq:etp}
   \hat{\text{N}}_{I(y)}^{\text{FC}} = \sum_{y'=1}^y \sum_{i'=1}^\text{i(y')-1} |\text{DoD}_{i'}^\text{cyc}| + \frac{|\text{DoD}_{i(y')}^\text{cyc}|}{2},
\end{equation}
where $I(y)$ is a set of segments used in a particular year $y$.

Finally, to linearize the product of binary and continuous variables in (\ref{eq:Pb_ch}) and (\ref{eq:Pb_dis}), the product of binary variables $\gamma_{1,c,I(1)}..\gamma_{y,c,I(y)} \theta_{y,t,k} = v_{I(y),k}$ has been linearized similar to the previous instance
\begin{equation}\label{eq:bin_prod2_lin}
    \begin{gathered}
    0 \leq v_{I(y),k} \leq 1,\\
    v_{I(y),k} \leq \gamma_{1,c,I(1)}, ...\\
    v_{I(y),k} \leq \gamma_{y,c,I(y)},\\
    v_{I(y),k} \leq \theta_{y,t,k},\\
    v_{I(y),k} \geq \gamma_{1,c,I(1)} + ... + \gamma_{y,c,I(y)} + \theta_{y,t,k} - y,
    \end{gathered}
\end{equation}
while the products of binary and continuous variables $v_{I(y),k} P_{y,t,g}^{\text{T}_\text{Ch}} = w_{I(y),k,g}$ and $v_{I(y),k} P_{y,t,h}^{\text{T}_\text{Dis}} = x_{I(y),k,h}$ have been liniarized as in (\ref{eq:bin_con_prod1_lin}) and (\ref{eq:bin_con_prod2_lin}), respectively
\begin{equation}\label{eq:bin_con_prod1_lin}
    \begin{gathered}
    w_{I(y),k,g} \leq |\text{P}_g^{\text{T}_\text{Ch}}| v_{I(y),k},\\
    P_{y,t,g}^{\text{T}_\text{Ch}} - |\text{P}_g^{\text{T}_\text{Ch}}|(1 - v_{I(y),k}) \leq w_{I(y),k,g} \leq P_{y,t,g}^{\text{T}_\text{Ch}},
    \end{gathered}
\end{equation}
\begin{equation}\label{eq:bin_con_prod2_lin}
    \begin{gathered}
    x_{I(y),k,h} \leq |\text{P}_h^{\text{T}_\text{Dis}}| v_{I(y),k},\\
    P_{y,t,h}^{\text{T}_\text{Dis}} - |\text{P}_h^{\text{T}_\text{Dis}}|(1 - v_{I(y),k}) \leq x_{I(y),k,h} \leq P_{y,t,h}^{\text{T}_\text{Dis}}.
    \end{gathered}
\end{equation}

\section{Numerical Study}\label{sec:num}
\subsection{Case Study}
\begin{figure}
\centering
        \includegraphics[width=0.75\textwidth]{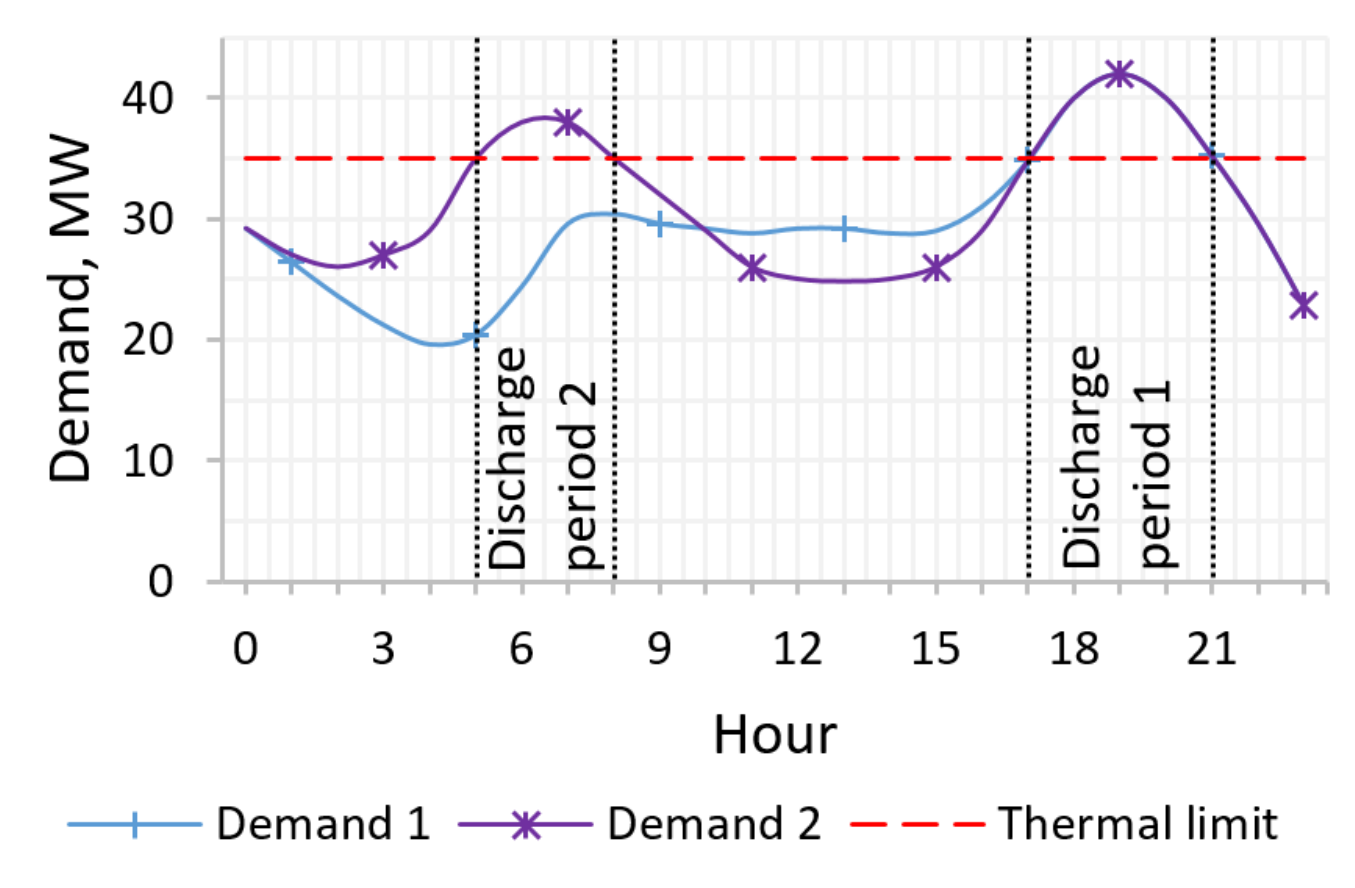}
        \caption{\textcolor{black}{Case study demand profiles}}
        \label{fig:demand}
\end{figure}

\textcolor{black}{This subsection describes a case study, using two realistic peak-shaving scenarios, depicted in Fig.~\ref{fig:demand}. Particularly,}  blue and purple curves represent demand profiles with one and two peaks, respectively. The red dashed line represents the maximum desired demand level. Both cases illustrate practically wide-spread scenarios, where the first case can correspond to a typical evening peak situation \cite{networkrevolution}, while the second - to a "duck curve" pattern due to massive photovoltaics integration \cite{iso2012duck}. 

\textcolor{black}{In the considered peak-shaving scenarios, battery storage is located downstream of the congestion to compensate for demand peaks that exceed the maximum desired level (e.g., thermal line limit or firm capacity). For instance, for one peak demand scenario, battery storage is required to compensate for a single peak by discharging from 17:00 to 21:00 (discharge period 1), where the instantaneous discharge power is determined by the difference between demand and the maximum demand level. Notably, battery charging can occur at any other time, where an optimizer determines particular charging power values (i.e., optimization problem variables). For two peak demand scenario, battery storage needs to compensate for two peaks during a day by discharging from 5:00 to 8:00 (discharge period 2) and from 17:00 to 21:00 (discharge period 1), where instantaneous discharge power, as well as charging time and charging power values are found similarly to the previous case. For both scenarios, the highest peak (discharge period 1) determines the minimum power and energy capacities required for peak-shaving. Particularly, the extent to which peak demand exceeds the maximum demand level determines the power capacity, while the total area between the maximum demand level and peak demand determines the energy capacity of an ideal lossless battery. For the considered case studies, these values correspond to 7 MW and 17.2 MWh, respectively.}

To focus on the optimal operation of the LiFePO$_4$ battery storage driven by its internal characteristics external factors \textcolor{black}{have been fixed} to constants, i.e., demand profiles remain unchanged during the battery lifetime and energy price $\text{C}^\text{En}$ is fixed to $80$ \$/MWh \cite{ofgem_inicators}. It is worth noting that the proposed approach allows considering variable energy price and a set of demand profiles for increasing load or stochastic problem formulation. Capital costs for battery power $\text{C}^\text{P}$ and energy $\text{C}^\text{E}$ capacities are $90$ \$/kW and $290$ \$/kWh, respectively \cite{fu20182018}. The End of Life (EoL) criterion is set to 75\%, while the planning horizon corresponds to the battery operational lifetime $\text{T}^\text{Lt}$, i.e., optimization problem variable.

\subsection{Results}

\begin{figure*}
\begin{center}
\subfigure[One peak demand]{\includegraphics[width=0.75\textwidth]{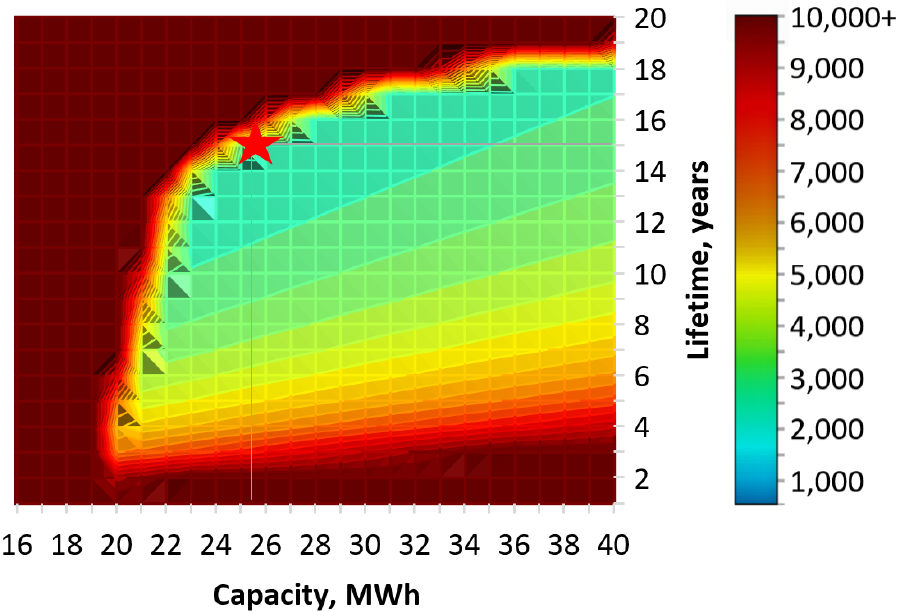}}
\hspace{0.01\textwidth}
\subfigure[Two peaks demand]{\includegraphics[width=0.75\textwidth]{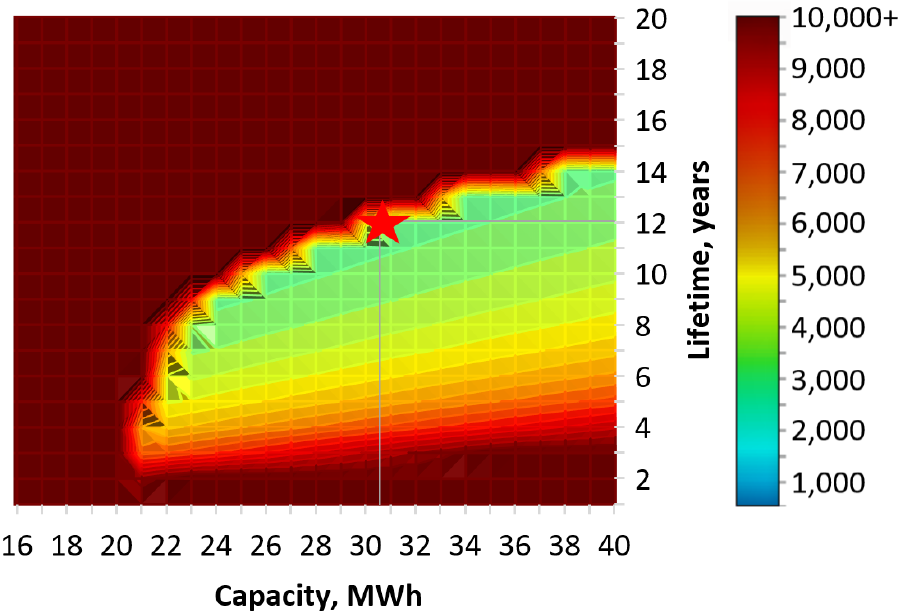}}
\caption{Objective function value as a function of installed energy capacity and operational lifetime}
\label{fig:Obj}
\end{center}
\end{figure*}

\begin{figure*}
\begin{center}
\textcolor{black}{\subfigure[One peak demand]{\includegraphics[width=0.75\textwidth]{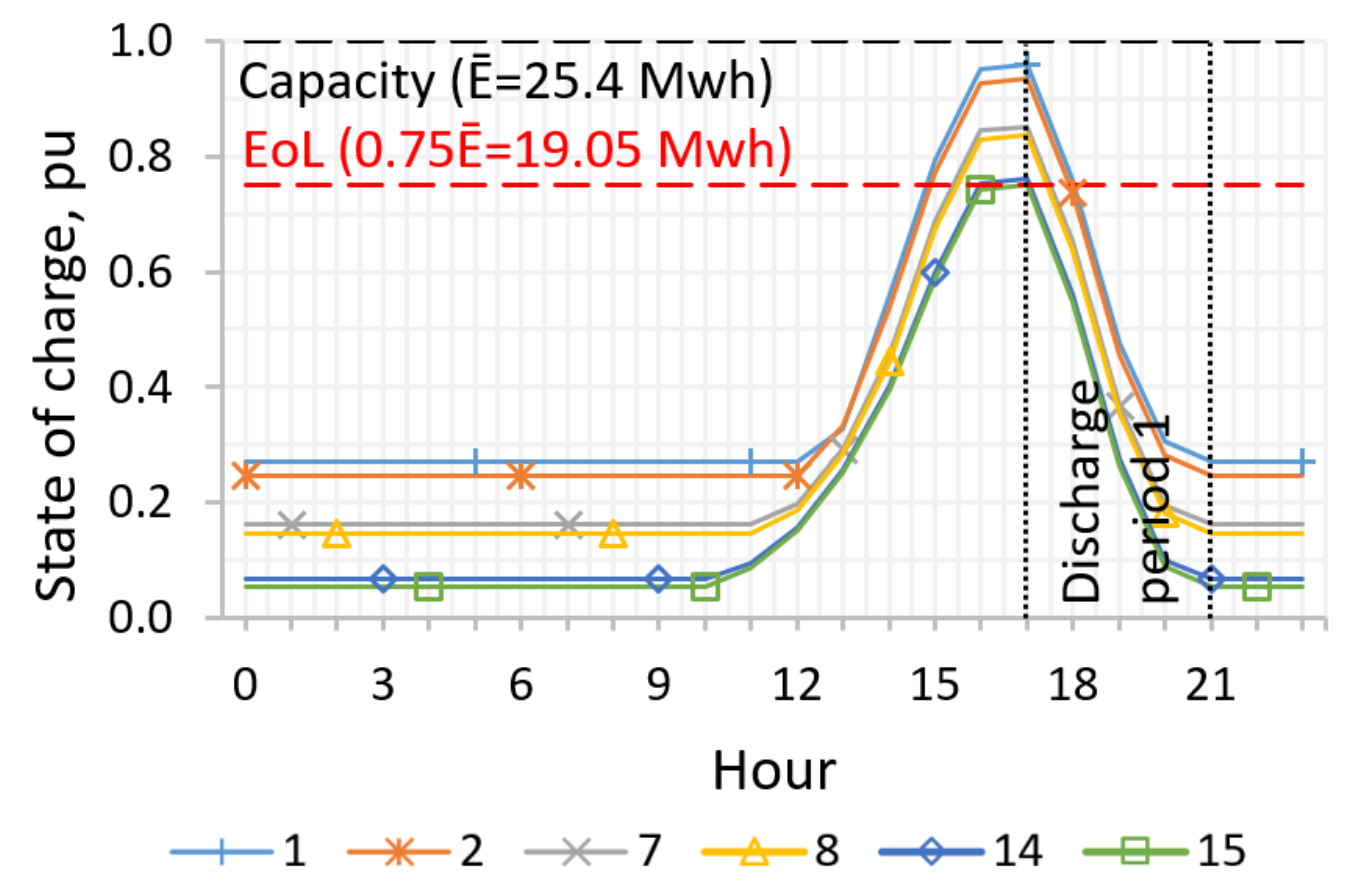}}}
\hspace{0.01\textwidth}
\textcolor{black}{\subfigure[Two peaks demand]{\includegraphics[width=0.75\textwidth]{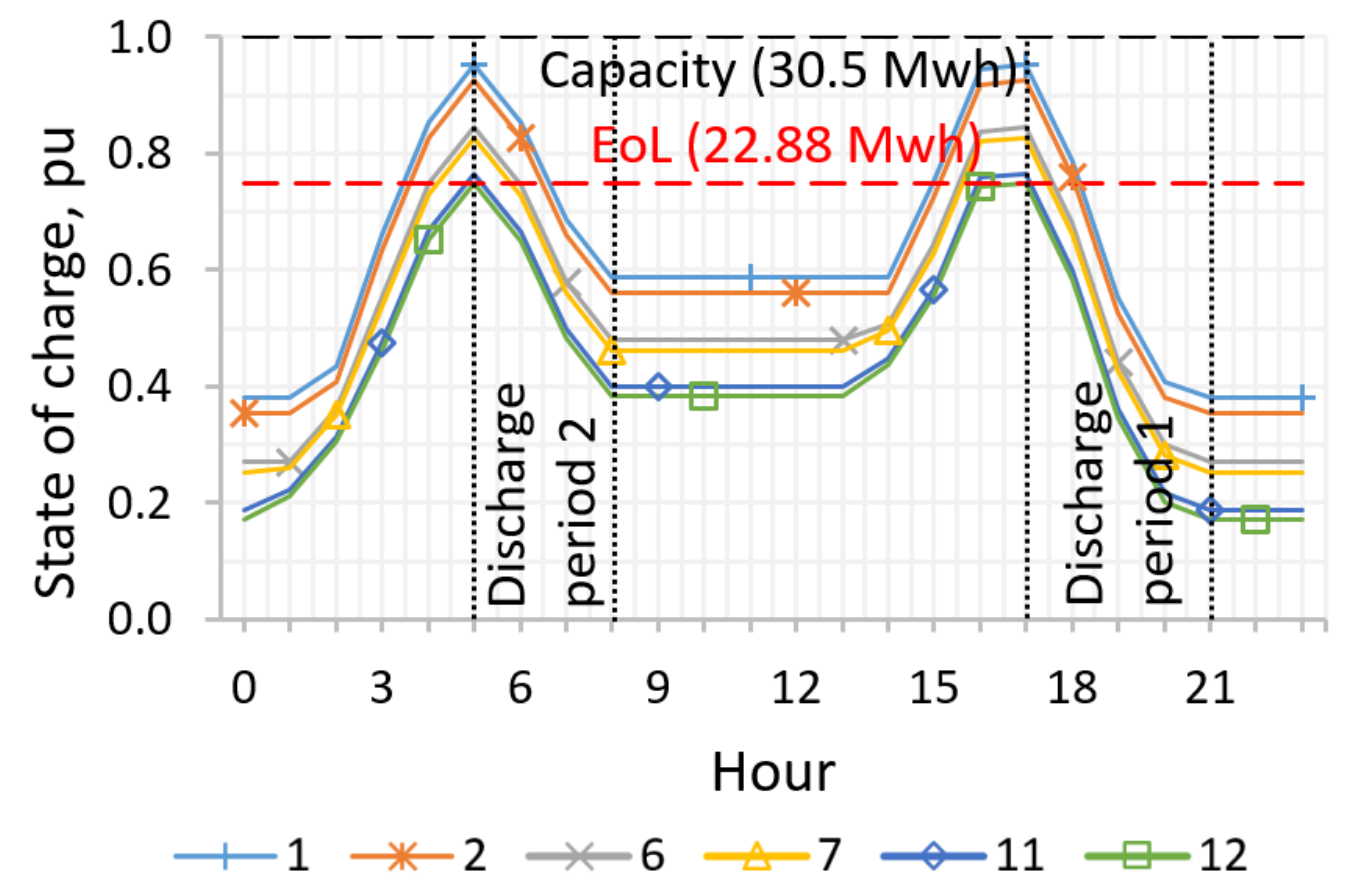}}}
\caption{\textcolor{black}{State of charge profiles of the optimal solution}}
\label{fig:SOC_res}
\end{center}
\end{figure*}

\begin{figure*}
\begin{center}
\subfigure[One peak demand]{\includegraphics[width=0.75\textwidth]{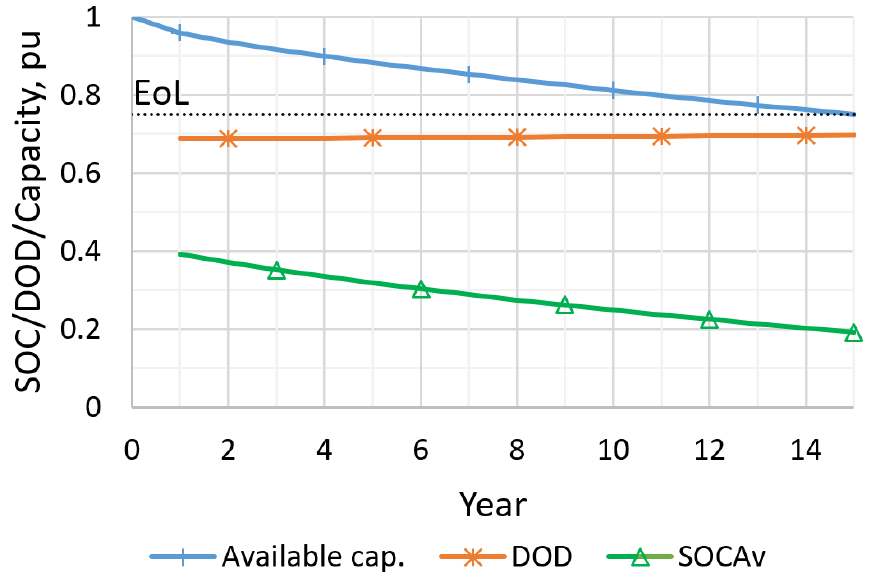}}
\hspace{0.01\textwidth}
\subfigure[Two peaks demand]{\includegraphics[width=0.75\textwidth]{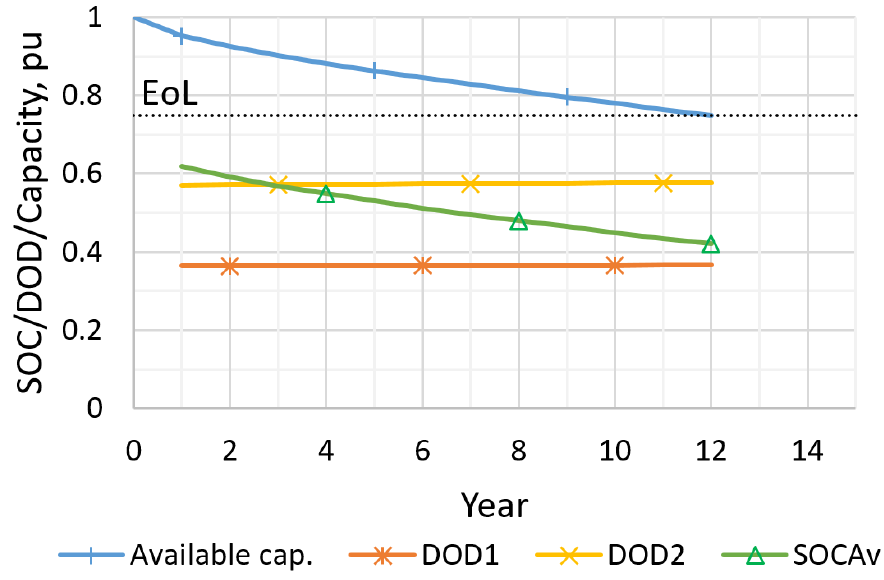}}
\caption{Operation and degradation characteristics of the optimal solution}
\label{fig:deg_res}
\end{center}
\end{figure*}

The main results of the formulated optimization problem applied to the case study above are provided in Table \ref{Table:results}. For the one peak demand scenario, the optimal solution corresponds to a $25.4$MWh/$7$MW battery system, which results in per diem battery investment and operating costs of $1512.1$\$/day for $15$ years of operational lifetime. For the two peak demand scenario, the optimal solution suggests installing $30.5$MWh/$7$MW battery storage, which corresponds to $2233.3$\$/day of per diem battery investment and operating costs for $12$ years of operation. Fig.~\ref{fig:Obj} illustrates the maps of the objective function in battery storage capacity $\bar{\text{E}}$ and operational lifetime $\text{T}^\text{Lt}$ space for two demand scenarios, while Fig.~\ref{fig:SOC_res} and Fig.~\ref{fig:deg_res} depict SoC, operation and degradation characteristics of the optimal solutions. Before analyzing the results, let us declare three major findings:

\begin{enumerate}
\item The optimal capacity of the LiFePO$_4$ battery is driven by the operating requirements, e.g., considerable capacity headroom becomes economically feasible for the case of two peaks per day. 
\item Given the gradient near the optimal solutions in Fig. \ref{fig:Obj}, it is safer to overestimate the capacity and underestimate operational lifetime than the opposite.
\item The operation strategy should be altered over the whole battery lifetime to ensure optimal utilization of the LiFePO$_4$ battery.
\end{enumerate}

\begin{table}[t]
\caption{Solutions of the optimization problem}\label{Table:results}
\begin{center}
\begin{tabular}{c | c | c | c}
\textbf{Demand} & \textbf{Objective} & \textbf{Capacity} & \textbf{Lifetime}\\ \hline
One peak & 1512.1 \$/day & 25.4 MWh / 7 MW & 15 years\\
Two peak & 2233.3 \$/day & 30.5 MWh / 7 MW & 12 years\\
\hline
\end{tabular}
\end{center}
\end{table}

As mentioned in the previous subsection, the minimum power and energy capacities to perform peak-shaving are $7$ MW and $17.2$ MWh, respectively. Even though the optimal solutions match the minimum power capacity requirement, there exists significant headroom in terms of energy capacity. For instance, the optimal battery energy capacities for one and two peak demand scenarios (Table \ref{Table:results}) correspond to $25.4$ MWh and $30.5$ MWh, which are 47.7\% and 77.3\% higher than the actual energy required to cover the highest peak, i.e., headroom. Even though the large part of these (33.3\%) corresponds to capacity fade compensation and around 2.5\% can be attributed to compensate for discharge losses, the remaining capacity margin is related to the operation strategy. Particularly, for the one peak demand scenario, this accounts for the remaining 11.9\% of energy capacity margin, while for the two peak demand scenario, where the battery is used more extensively, this accounts for the remaining 41.5\% of headroom to achieve optimal utilization of the LiFePO$_4$ battery storage. A similar effect is observed in the results of \cite{berrueta2018combined}, where the authors provide qualitative analysis of the reasons behind the excessive headroom capacity obtained from optimization. Particularly, the authors refer to the lower discharge current and smaller SoC range, which reduce the battery aging.

In contrast to the above solutions, a naive strategy would be to choose the battery capacity accounting only for the minimum required energy capacity, EoL criterion, and discharge efficiency, e.g., $17.2$ MWh $/0.75/0.98 = 23.4$ MWh. Even though the derived battery capacity would require less capital investments, compared to the obtained solutions, the resulting per diem investment and operating costs would be higher due to shorter operational lifetime (11 and 8 years for one peak and two peaks demand scenarios, respectively).


Fig. \ref{fig:Obj} illustrates the positions of the optimal solutions in the objective function value map, which is presented as a function of installed energy capacity and operational lifetime. The red stars indicate the minimum objective function value positions, i.e., the optimal solutions. For the one peak demand scenario (a), the minimum objective function value equals $1512.1$ \$/day, which corresponds to $25.4$ MWh of installed energy capacity and $15$ years of battery lifetime. For the two peak demand scenario (b), the minimum objective function value is found at the intersection of $30.5$ MWh of installed energy capacity and $12$ years of operational lifetime and equals $2233.3$ \$/day. As it can be seen from Fig. \ref{fig:Obj}, both solutions are located very close to the high gradient of the objective function, meaning that the minor disturbance (error) of the optimal solution might result in a significant increase in the objective function value. Particularly, the profitability of a solution might be significantly compromised if the capacity is underestimated and operational lifetime is overestimated. However, one might want to overestimate the installed energy capacity and underestimate operational lifetime to reduce the sensitivity and investment risks at the cost of a minor increase in investment and operating costs.

Fig. \ref{fig:SOC_res} illustrates the optimal LiFePO$_4$ battery scheduling during the whole operational lifetime period. In the case of the one peak demand scenario (a), the SoC profile variation changes from [27\%;95.8\%] range at the beginning of the battery lifetime to [5.4\%;75\%] during the terminal year. A similar picture is observed for the case of the two peak demand scenario (b), where the SoC ranges of two consecutive cycles change from [58.8\%;95.2\%] and [38\%;95.2\%] during the first year of operation to [38.4\%;75\%] and [17.3\%;75\%] during the terminal year, respectively. Even though the span of the ranges, i.e., DoD, increases only by 0.8\% for the one peak demand scenario (a) and by 0.2\% and 0.5\% for the two consecutive peaks of the two peak demand scenario (b), the battery SoC strategy changes through the whole lifetime period quite significantly. For instance, the gradual decrease of the average battery SoC can be observed in Fig. \ref{fig:deg_res}, where in the case of the one peak demand scenario (a) it drops from 39.3\% to 19.1\%, and in the case of the two peak demand scenario (b), it falls from 61.8\% to 42.1\%. Since the DoD is tied to the amount of energy required to shave the peak, it cannot be changed once the battery capacity is chosen. Thus, the only operation characteristic that can be altered is the SoC, which is observed in the numerical study.

Given the constant peak-shaving requirements for the entire battery lifetime period, the small increase in the DoD strategy is explained by the need to compensate for the increased discharge losses associated with the internal resistance growth. The substantial alternation of the battery operation strategy relates to both internal resistance growth and capacity fade characteristics. As per (\ref{eq:cap_fade_idl}) and (\ref{eq:cap_fade_cyc}), the battery SoC is in direct relation to the capacity fade from idling, while the median cycle SoC is in inverse relation to the capacity fade from cycling. Thus, in Fig. \ref{fig:SOC_res}, a rapid charge of the battery \textcolor{black}{is observed} just before it is required to discharge. This way, it is possible to keep the average battery SoC low while the median cycle SoC is high, which is complementary to the slow degradation process. However, given the fact that the average daily SoC decreases asymptotically with the available capacity (see Fig. \ref{fig:deg_res}), it can be concluded that capacity fade from cycling is the dominating factor. Also, it can be noted that during the course of the battery lifetime, the time duration for battery charging is increased from four hours at the beginning of the battery lifetime to seven hours during the terminal year (see Fig. \ref{fig:SOC_res}), which negatively affects the average daily SoC. This reflects the time-varying trade-off between power losses and capacity fade from idling, where the latter dominates during the early battery lifetime, while the former comes to the fore after.

\subsection{Comparative analysis}

To quantify the advantages of the proposed modelling approach, it has been compared to two existing battery sizing methodologies. The first methodology (referred to as "Cyc.Lt.(DoD,C-rate)") is taken from \cite{padmanabhan2019battery}, where the nonlinear relationship between the battery DoD, C-rate, and cycle lifetime is considered with a piece-wise linear function. However, in contrast to the proposed methodology, the battery efficiency and available battery capacity are kept constant. The second methodology (referred to as "Deg.(SoC,DoD,C-rate);Rint(SoC)") has been proposed in \cite{berrueta2018combined}, where the dynamic programming optimization is used to resolve a comprehensive Li-ion battery model that accounts for the battery degradation (i.e., capacity fade and internal resistance growth from both idling and cycling), and SoC dependant equivalent circuit $Rint$ model. In contrast to the proposed approach, both methodologies allow choosing only one battery operation strategy for the whole planning horizon, while, as it has been shown in the previous subsection, to achieve optimal battery utilization, the strategy needs to be substantially altered during the whole operational lifetime (see Fig. \ref{fig:SOC_res} and \ref{fig:deg_res}). Notably, all three methodologies have been applied to the same LiFePO$_4$ benchmark model from the literature and the same case study of one peak demand scenario, described in the present section. The first state-of-the-art model from \cite{padmanabhan2019battery} is MILP compatible, and it was solved using the same solvers used for the proposed model. The second state-of-the-art model from \cite{berrueta2018combined} does not meet MILP requirements; hence, it was solved using a dynamic programming heuristic proposed in the same paper. It is worth noting that given the same disposition of the sizing methodologies to possible errors (investment risks), the obtained solutions would be indicative for the relative expected benefit of one method over the other if the original model is the same. Thus, the advantage of the proposed methodology over the state-of-the-art \textcolor{black}{is derived} based on the obtained optimal solutions.

\begin{table}[t]
\caption{Comparative study}\label{Table:compare}
\begin{center}
\begin{tabular}{c | c | c | c}
\textbf{Model} & \textbf{Objective} & \textbf{Capacity} & \textbf{Lifetime}\\ \hline
Cyc.Lt.(DoD,C-rate)\cite{padmanabhan2019battery}&1879.5\$/day&23.4MWh/7MW&11years\\
Deg.(SoC,DoD,C-rate);Rint(SoC)\cite{berrueta2018combined}&1695.5\$/day&29.0MWh/7MW&15years\\
Proposed&1512.1\$/day&25.4MWh/7MW&15years\\
\hline
\end{tabular}
\end{center}
\end{table}

\begin{figure}
\centering
        \includegraphics[width=0.75\textwidth]{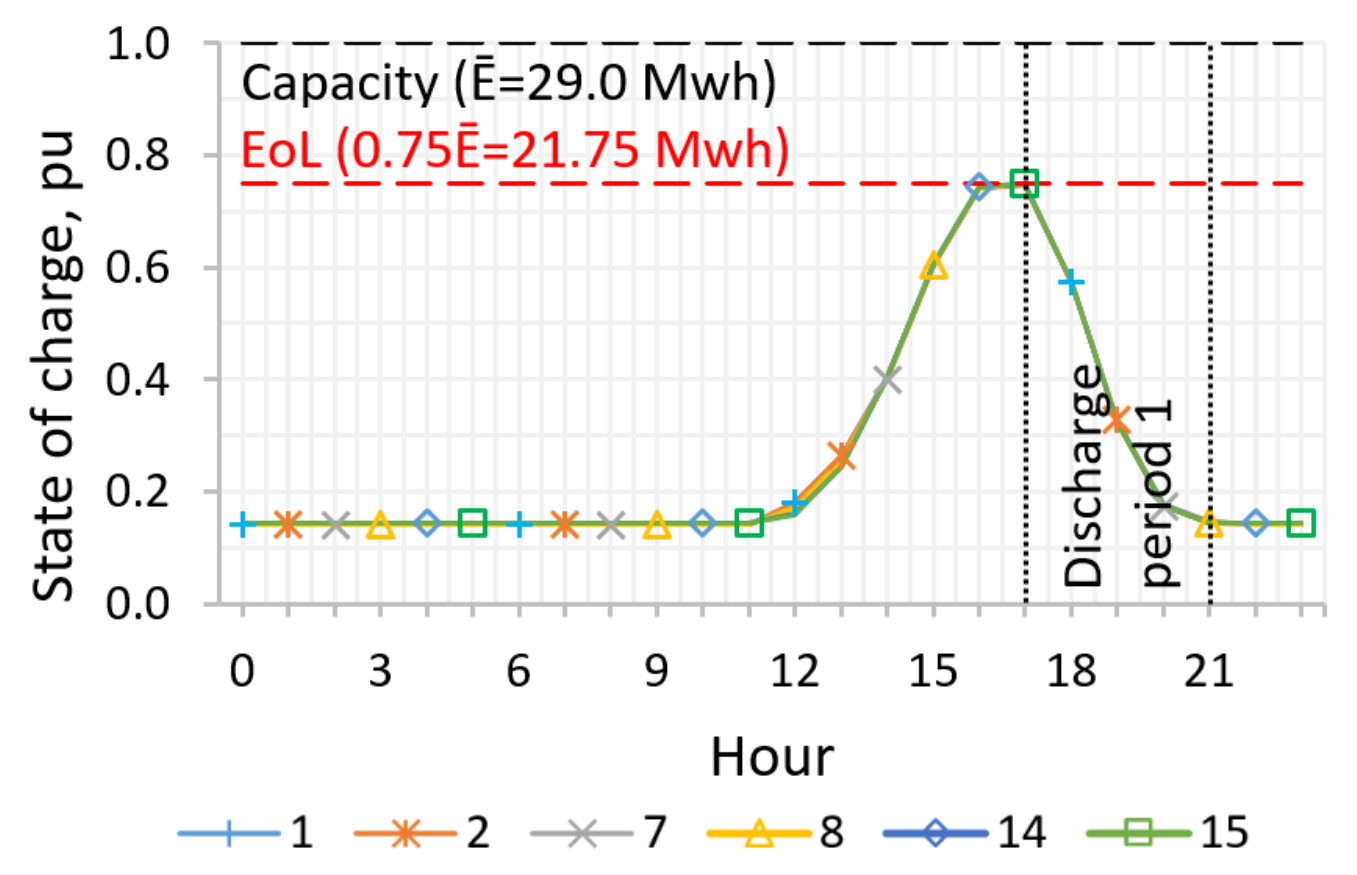}
        \caption{\textcolor{black}{Deg.(SoC,DoD,C-rate);Rint(SoC) \cite{berrueta2018combined} solution - State of charge profile}}
        \label{fig:SOC_els}
\end{figure}

The results of the three approaches under comparison are given in Table \ref{Table:compare}. In the case of the variable battery lifecycle (Cyc.Lt.(DoD,C-rate)), the solution suggests installing a $23.4$MWh/$7$MW battery system, which results in daily investment and operating costs of 1879.5 \$/day. The optimal DoD is found to be 75\%, which corresponds to the EoL criterion and leads to 4,000 cycles or 11 years. In case of the comprehensive battery modelling approach (Deg.(SoC,DoD,C-rate);Rint(SoC)), the optimal solution suggests installing $29.0$MWh/$7$MW battery system, which results in the objective function value of 1695.5 \$/day. The solution corresponds to the battery dispatch depicted in Fig.~\ref{fig:SOC_els}, where the operation strategy is found to be 25.5\% average battery SoC, 44.7\% cycle median SoC, and 60.7\% cycle DoD over the whole battery lifetime, which in this case is found to be $15$ years. In its turn, the optimal solution obtained by the proposed approach corresponds to $25.4$MWh/$7$MW battery system, which corresponds to the objective function value of 1512.1 \$/day for 15 years of expected battery lifetime. As per Fig. \ref{fig:SOC_res} (a) and \ref{fig:deg_res} (a), the optimal battery utilization corresponds to the operation characteristics that evolve through the whole lifetime period. Particularly, the average battery SoC changes from 39.3\% at the beginning of the battery lifetime to 19.1\% during the terminal year, cycle median SoC changes from 61.4\% to 40.2\%, and cycle DoD changes from 68.8\% to 69.6\%. Compared to the previous approach, such an adjustable operation strategy allows providing the same service for the same planning horizon with a substantially smaller battery capacity. Particularly, the battery energy capacity found by the approach in \cite{berrueta2018combined} is 14.2\% higher than the one found by the proposed method, what leads to a 12.1\% reduction of the objective function value, i.e., investment and operating costs.

\section{Discussions}\label{sec:dis}

\subsection{Li-ion battery model relationship with the MILP problem formulation}
\begin{figure}
\centering
        \includegraphics[width=1\textwidth]{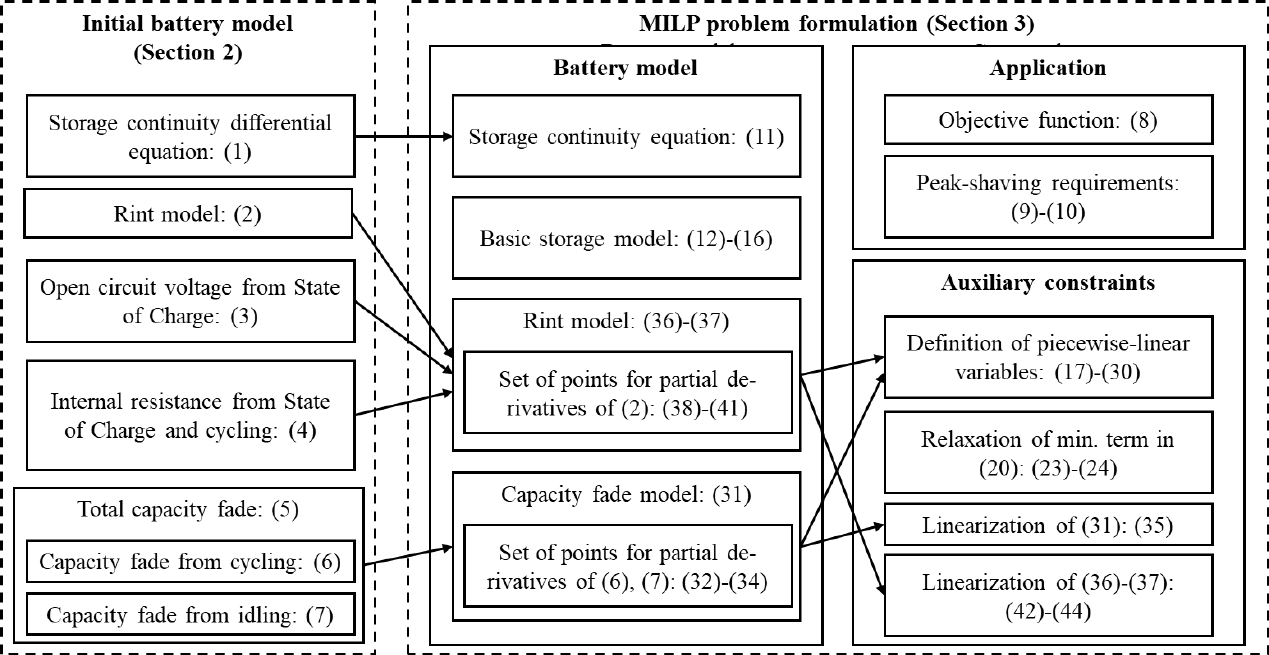}
        \caption{Li-ion battery model relationship with the MILP problem formulation}
        \label{fig:model_rel}
\end{figure}

Fig. \ref{fig:model_rel} illustrates the relationships of the considered in Section \ref{sec:model} models with the proposed problem formulation in Section \ref{sec:case_study}. The objective function \eqref{eq:objective} and constraints \eqref{eq:balance}-\eqref{eq:thermal_limit} constitute the battery application of the proposed problem formulation, which are system-dependent with the initial battery model. Storage continuity equation \eqref{eq:SOC} is derived from the storage dynamic equation \eqref{eq:charge_diff}, while the constraints \eqref{eq:net_charge}-\eqref{eq:eol_cond} formulate a basic storage model, which is widely used in the literature. The constraints \eqref{eq:Pb_ch}-\eqref{eq:Pb_dis} incorporate Rint model \eqref{eq:Pb}, open-circuit voltage model \eqref{eq:Voc}, and internal resistance model \eqref{eq:Rin}. The auxiliary constraints \eqref{eq:pt_ch}, \eqref{eq:pt_dis}, \eqref{eq:soc_cur}, \eqref{eq:pt_ch_aux}, \eqref{eq:pt_dis_aux}, \eqref{eq:soc_cur_aux}, and \eqref{eq:Pb_ch_aux1}-\eqref{eq:bin_con_prod2_lin} are formulated for piecewise linear representation of the initially nonlinear model. The constraint \eqref{eq:cap_fade_lin} constitutes the model \eqref{eq:cap_fade}, which consists of the capacity fade models from idling \eqref{eq:cap_fade_idl} and cycling \eqref{eq:cap_fade_cyc}. The auxiliary constraints \eqref{eq:dod_cyc}-\eqref{eq:soc_idl}, \eqref{eq:e_min},  \eqref{eq:e_min_con}, \eqref{eq:dod_cyc_aux}-\eqref{eq:soc_idl_aux}, \eqref{eq:dod_cyc_str}-\eqref{eq:bin_prod1_lin} are used for piecewise linear representation of the initially nonlinear model.

\subsection{Vehicular applications}
The proposed approach allows considering vehicular applications in a similar manner it is done for the grid peak-shaving. Let us consider a general two peak demand scenario from the case study. The peak shaving requirements, which are determined by the curve above the power line thermal limit (dashed line in Fig. \ref{fig:demand}), can illustrate a daily driving pattern of an electric vehicle. The first peak represents a trip from home to work, while the second peak represents a return trip with a small deviation from a route to visit a supermarket. Acknowledging that for power systems and vehicular applications, discharge power requirements differ in magnitude and in time, the charging schedule for both applications is a subject of optimization. Moreover, in some cases, the battery capacity may need to be optimized as well. Thus, \textcolor{black}{there are} at least two possibilities of the proposed problem formulation to be used for vehicular applications. First, it can be used within the onboard energy management system to determine the optimal charging strategy to minimize battery degradation and power losses. Second, it can be used by EV design engineers to determine the optimal battery capacity to prolong the expected battery lifetime while keeping the cost for battery optimal.

\subsection{Temperature effect}
\textcolor{black}{The Li-ion battery performance is greatly affected by the temperature. According to \cite{ji2013li}, the acceptable temperature range of the Li-ion batteries is from -20\degree C to 60\degree C, while the desired range is within 15\degree C and 35\degree C \cite{pesaran2013addressing}. The desired range ensures that the battery performance is high while degradation is moderate. To manage battery temperature, various thermal management systems are used in practice. For instance, to keep batteries from overheating, the most effective solutions include air cooling, liquid cooling, and phase change material cooling, while to keep batteries warm, they can be equipped with climate control systems that are more suitable for stationary batteries \cite{ma2018temperature}. For the reasons above and to focus on other aspects of the Li-ion battery applications, authors in the scientific literature usually assume constant battery temperature. Such that, in both state-of-the-art methods described in the literature review, constant temperature values are considered. For instance, authors in \cite{li2020design} fix the battery temperature at 25\degree C, while in \cite{berrueta2018combined} battery temperature value is set to 30\degree C. Similar to the state-of-the-art,} in the present study, the battery cell temperature was considered constant at 25\degree C.

\textcolor{black}{In addition to the above,} most of the battery models found in the literature are formulated for a specific value of temperature, and they do not provide enough information to represent the dependencies from temperature using standard differentiable functions. 
However, if the dependencies from the temperature were known for all relevant battery models, it would be possible to account for them within the proposed problem formulation. In this case, the battery temperature evolution could be considered with additional constraints that formulate a thermal model of a battery system. \textcolor{black}{For reference}, a linear battery system thermal model can be found in \cite{sayfutdinov2020alternating}. Finally, the nonlinear characteristics from temperature can be formulated with piecewise-linear approximation using \emph{Special Order Sets 2}, as it was done for DoD and SoC in the degradation and other models.

\subsection{Software and hardware}
The proposed MILP optimization problem has been formulated in \textcolor{black}{Julia for Mathematical Programming (JuMP)}. The Ipopt and Juniper solvers have been used for the linear and mixed-integer parts of the optimization problem. The former exploits the interior-point algorithm for the linear relaxations of the proposed MILP problem, while the latter uses the Branch-and-Bound algorithm to manage the mixed-integer part. The tolerance of both solvers has been set to 1e-6, while the other parameters remained in default. The optimization problems for two case studies have been solved on Intel Core i5-7200 CPU @ 2.50 GHz and 2.71 GHz, 8 GB RAM laptop computer. The computational time required to solve the problems did not exceed 600 seconds.

\section{Conclusions}\label{sec:con}

To sum up, the contribution of the present paper is three-fold. Firstly, a MILP compatible comprehensive battery modelling approach is developed, which accounts for numerous Li-ion battery characteristics, i.e., degradation from idling and cycling, internal resistance as a function of both degradation and SoC, as well as the equivalent circuit model to account for nonconstant battery efficiency. The nonlinear characteristics have been linearized using the \emph{Special Order Sets 2} to be suitable for use within the MILP problems, e.g., optimal scheduling and sizing. Secondly, the distinctive advantage of the proposed methodology, compared to the existing approaches, resides in the fact that the operation strategy of a battery storage system can be optimized for each lifetime period separately, i.e., separate variables of the optimization problem. In the comparative study, it has been shown that the particular value associated with such an adaptable operation strategy accounts for 12.1\% of the total battery system budget, i.e., investment and operation costs. Finally, applying the developed LiFePO$_4$ battery model to realistic case studies, it has been found that the optimal utilization of the battery corresponds to the varying over the lifetime operation strategy. This includes increasing DoD to compensate for the growing internal resistance and associated charge and discharge losses, decreasing median cycle SoC to minimize battery degradation from cycling, and increasing average SoC and battery charging process duration as a trade-off between degradation from idling and growing charge and discharge losses. \textcolor{black}{In future work, it is planned to include the temperature effect into the proposed modelling approach to study the impact of different regions and ambient temperatures on the battery utilization patterns.}
\section{Acknowledgments}\label{sec:ack}

Work of T.~Sayfutdinov was supported by Skoltech and Newcastle University.
Work of P.~Vorobev was Supported by Skoltech and The Ministry of Science and Higher Education of Russian Federation, Grant~29 Agreement No. 075-10-2021-067, Grant identification code 30 000000S707521QJX0002.






\bibliographystyle{elsarticle-num-names}
\bibliography{sample.bib}







\end{document}